# Efficient Calculations with Multisite Local Orbitals in a large-scale DFT Code CONQUEST

*Ayako Nakata,*[*,†,‡] *David R. Bowler,*[§,‖,⊥] *and Tsuyoshi Miyazaki*[**,‡,§]

[†]International Center for Young Scientists (ICYS), National Institute for Materials Science (NIMS), 1-1 Namiki, Tsukuba, Ibaraki 305-0044, Japan

[‡]Computational Materials Science Unit (CMSU), National Institute for Materials Science (NIMS), 1-1 Namiki, Tsukuba, Ibaraki 305-0044, Japan

[§]Department of Physics & Astronomy, University College London, Gower St, London WC1E 6BT, UK

[‖]WPI-MANA, National Institute for Materials Science (NIMS), 1-1 Namiki, Tsukuba, Ibaraki 305-0044, Japan

[⊥]London Centre for Nanotechnology, University College London, 17-19 Gordon Street, London WC1H 0AH, UK

ABSTRACT

Multisite local orbitals, which are formed from linear combinations of pseudo-atomic orbitals from a target atom and its neighbor atoms, have been introduced in the large-scale density functional theory calculation code CONQUEST. Multisite local orbitals correspond to local molecular orbitals so that the number of required local orbitals can be minimal. The multisite support functions are determined by using the localized filter diagonalization (LFD) method [Phys. Rev. B 2009, 80, 205104]. Two new methods, the double cutoff method and the smoothing method, are introduced to the LFD method to improve efficiency and stability. The Hamiltonian and overlap matrices with multisite local orbitals are constructed by efficient sparse-matrix multiplications in CONQUEST. The investigation of the calculated energetic and geometrical properties and band structures of bulk Si, Al and DNA systems demonstrate the accuracy and the computational efficiency of the present method. The representability of both occupied and unoccupied band structures with the present method has been also confirmed.

1. INTRODUCTION

Density functional theory (DFT) calculations have been applied widely on both condensed phase materials and molecules. Plane-wave basis functions have been often used for condensed phase materials, while atomic orbital basis functions have been widely used for molecules. Plane-wave bases show systematic convergence of the accuracy as the cutoff energy increases, but calculations with large cutoff energies have a huge computational cost. On the other hand, atomic orbital basis can provide reasonable accuracy with small number of basis functions, but systematic improvement with respect to the number of the functions is difficult.

We have been developing a linear-scaling DFT code CONQUEST using real-space basis functions with periodic-boundary condition.[1-3] The computational cost of linear-scaling DFT is proportional to the number of atoms in the target system $N$, while the conventional DFT calculations scale cubically to $N$. 'Support functions' are used to express the density matrix in CONQUEST. The support functions are localised orbitals which are linear combinations of the given basis functions and are carefully optimized for each target system. The locality of support functions plays a significant role to reduce the computational cost in CONQUEST, because Hamiltonian, overlap, and density matrices are constructed as sparse. CONQUEST provides two types of basis functions to express the support functions: b-spline (blip) finite-element functions[4] akin to plane-waves; and pseudo atomic orbitals (PAOs)[5]. We focus on the use of PAOs in the present study.

PAOs are confined atomic orbital basis which are used with pseudopotentials and consist of radial functions multiplied by spherical harmonic functions.[6,7] Although it is usually difficult to achieve systematic convergence, the accuracy of the calculations with PAOs can be improved by increasing the number of PAOs on each atom. When increasing the number of PAOs, we often prepare 'multiple-$\zeta$' functions in which several radial functions are combined with a spherical harmonic function. There are two ways to construct support functions with multiple-$\zeta$ PAOs[5]: one is to use each PAO as a single support function; the other is to express a support function as a linear combination of multiple PAOs. These cases are called 'primitive' and 'contracted' support functions. The primitive support functions have no flexibility but provide the most accurate results possible with the given PAO basis. As for the contracted support functions, we need to optimize the total energy with respect to the linear-combination coefficients, but the number of support functions becomes smaller than for the primitive case. The idea of the contraction of atomic orbital basis functions is popular in quantum chemistry calculations,[8] in which the linear-combination coefficients are usually fixed to the values optimized for some representative molecules. On the other hand, the coefficients in the contracted support functions in CONQUEST are optimized for each atom in each target system. Ozaki and Kino have also proposed a similar method to optimize the coefficients and Kohn-Sham density simultaneously,[9,10] though in practice the coefficients are often fixed, as this optimization procedure requires large computational time. Moreover, since they are constructed only from PAOs which are centered at the target atom, the conventional support functions have to keep the point-group symmetry of the target atom.[5] This constraint leads a limitation in reducing the number of support functions, which will have an impact on the computational cost. In CONQUEST, the most expensive operations are the multiplication of sparse matrices in linear scaling and the diagonalization of Hamiltonian in exact diagonalization. In both cases, the cost increases cubically with the number of support functions on each atom. Therefore, it is crucial to reduce the number of support functions for the efficiency, while keeping as much accuracy as possible.

In this study, we introduce 'multisite' support functions, to reduce the number of the contracted support functions by increasing their flexibility. Multisite support functions are constructed from not only the PAOs on the target atom but also the PAOs of its neighbor atoms, which makes the support functions free from the atomic-symmetry constraint. The quality of the linear-combination coefficients in multisite support functions is important to keep the accuracy comparable to the primitive support functions. To determine the coefficients, we apply the localized filter diagonalization (LFD) method[11, 12] which has been proposed by Rayson and Briddon recently, and in which the coefficients are determined by projecting the local molecular orbitals (MOs) to the PAOs in a region around the target atom. Lin et al have also proposed a similar method to construct multisite orbitals from a kind of finite-element basis functions.[13, 14] We introduce the double cutoffs to the LFD method: one cutoff to calculate the local MOs, and another cutoff for the locality of the support functions. We also introduce a smoothing function to avoid the abrupt change in the accuracy of multisite support functions, which comes from the change of the number of neighbor atoms. With these techniques, we will show that the method using multisite support functions is a promising tool for robust, accurate and efficient large-scale DFT calculations with CONQUEST.

The rest of the paper is organized as follows: in the next section we explain the method of multisite support functions; the performance of the multisite support functions are examined for bulk silicon (Si), aluminum (Al) and DNA systems in the third section; the final section gives the conclusion of the present study.

## 2. METHODOLOGY

### 2.1. Multisite Support Functions in CONQUEST

The Kohn-Sham density matrix $\rho$ in DFT is defined as

$$\rho(\mathbf{r},\mathbf{r}') = \sum_n f_n \psi_n(\mathbf{r}) \psi_n(\mathbf{r}')^*, \qquad (1)$$

where $\psi_n$ is $n$th Kohn-Sham eigenfunction and $f_n$ is its occupation number. In CONQUEST, $\rho$ is rewritten in the support function basis[1] as

$$\rho(\mathbf{r},\mathbf{r}') = \sum_{i\alpha, j\beta} \phi_{i\alpha}(\mathbf{r}) K_{i\alpha,j\beta} \phi_{j\beta}(\mathbf{r}')^*. \qquad (2)$$

Indices $i$ and $j$ correspond to atoms, and $\alpha$ and $\beta$ to support functions. $K$ is the density matrix in support function basis. PAOs with one, two, and three radial functions for valence orbitals are called single-$\zeta$ (SZ), double-$\zeta$ (DZ), and triple-$\zeta$ (TZ) PAOs, respectively. Additional functions are often adopted to describe polarization (P), for example, triple-$\zeta$ with polarization (TZP). A conventional support function consisting of multiple-$\zeta$ PAOs[5] is written as

$$\phi_{i\alpha}(\mathbf{r}) = \sum_\mu c_{i\alpha,i\mu} \chi_{i\mu}(\mathbf{r}). \qquad (3)$$

where $\chi_{i\mu}$ is $\mu$th PAO on atom $i$. 'Primitive' support functions are equivalent to the given PAOs without any modification. In this case, the linear-combination coefficients $c$ in Eq. (3) are always 1 in the primitive support functions. In our previous works, contracted support functions for atom $i$ had consisted of the PAOs which belong to $i$ as in Eq. (3). We refer them to 'single-site' support functions. The coefficients $c$ in Eq. (3), which we call PAO coefficients of the support function, are optimized with a minimization method, such as the conjugate-gradient method, using the gradient of the total energy with respect to the PAOs.

Here we introduce the 'multisite' support function which is constructed as a linear combination of the PAOs not only on the target atom but also on neighbouring atoms:

$$\phi_{i\alpha}(\mathbf{r}) = \sum_{k}^{neighbors} \sum_{\mu \in k} C_{i\alpha,k\mu} \chi_{k\mu}(\mathbf{r}). \tag{4}$$

In Eq. (4) the summation of atom $k$ runs over neighbouring atoms which are within the range $r_{MS}$ of the target atom $i$, including atom $i$ itself. Since these support functions have more degrees of freedom than single-site support functions, and are free from the constraint of the symmetry of atomic orbitals, the number of multisite support functions can be smaller than that of single-site support functions. The coefficient matrix $\mathbf{C}$ is sparse because only the atoms within $r_{MS}$ are taken into account as neighbor atoms. The overlap matrix of multisite support functions is

$$S_{i\alpha,j\beta} = \int \phi_{i\alpha} \phi_{j\beta}^* dr = \sum_{k\mu} \sum_{p\nu} C_{i\alpha,k\mu} \langle \chi_{k\mu} | \chi_{p\nu} \rangle C_{p\nu,j\beta}^* = \sum_{k\mu} \sum_{p\nu} C_{i\alpha,k\mu} S_{k\mu,p\nu}^{PAO} C_{p\nu,j\beta}^* \tag{5}$$

$\chi_{p\nu}$ is the $\nu$th PAO on atom $p$ which is a neighbor of atom $j$ within $r_{MS}$. Since each PAO is completely localized in a finite region with the cutoff range $r_{PAO}$, the overlap matrix $\mathbf{S}^{PAO}$ between PAOs has nonzero values only if the distance between the two atoms ($k$ and $p$) is smaller than $2r_{PAO}$. Therefore, the calculations of Eq. (5) can be performed efficiently using the technique of parallel sparse matrix multiplications implemented in CONQUEST,[15]

$$\mathbf{S} = \mathbf{C}\mathbf{S}^{PAO}\mathbf{C}^\dagger. \tag{6}$$

The range of $\mathbf{S}$ is equal to the twice of ($r_{MS} + r_{PAO}$).

In the evaluation of the Hamiltonian matrix, the Hartree and exchange-correlation potentials are calculated from the electron density on regular grids using FFT. The electron density on each grid point is obtained from multisite support functions as

$$n(\mathbf{r}_g) = \sum_{i\alpha,j\beta} \phi_{i\alpha}(\mathbf{r}_g) K_{i\alpha,j\beta} \phi_{j\beta}(\mathbf{r}_g) = \sum_{i\alpha,j\beta} \sum_{k\mu} \sum_{p\nu} C_{i\alpha,k\mu} \chi_{k\mu}(\mathbf{r}_g) K_{i\alpha,j\beta} C_{j\beta,p\nu} \chi_{p\nu}(\mathbf{r}_g). \qquad (7)$$

Given the charge density, we can calculate the Hamiltonian matrix in the PAO basis $\mathbf{H}^{\text{PAO}}$, whose matrix elements are nonzero only when the distance between the atoms of two PAOs is smaller than $2r_{\text{PAO}}$.[16] Then, we can calculate the Hamiltonian matrix expressed with the multisite support functions $\mathbf{H}$ from $\mathbf{H}^{\text{PAO}}$ as

$$\mathbf{H} = \mathbf{C}\mathbf{H}^{\text{PAO}}\mathbf{C}^\dagger. \qquad (8)$$

The cutoff of the matrix $\mathbf{H}$ is the same as $\mathbf{S}$, $2(r_{\text{MS}} + r_{\text{PAO}})$. From $\mathbf{H}$ and $\mathbf{S}$, we can obtain the density matrix $\mathbf{K}$ using a linear-scaling method or a diagonalization method. Furthermore, if we introduce the density matrix $\mathbf{K}^{\text{PAO}}$ in the PAO basis defined as

$$K^{\text{PAO}}_{k\mu,p\nu} = \sum_{i\alpha,j\beta} C^*_{k\mu,i\alpha} K_{i\alpha,j\beta} C_{j\beta,p\nu} \quad \Rightarrow \quad \mathbf{K}^{\text{PAO}} = \mathbf{C}^\dagger \mathbf{K} \mathbf{C}, \qquad (9)$$

the electronic density can be written in PAO basis:

$$n(\mathbf{r}_g) = \sum_{k\mu} \sum_{p\nu} \chi_{k\mu}(\mathbf{r}_g) K^{\text{PAO}}_{k\mu,p\nu} \chi_{p\nu}(\mathbf{r}_g). \qquad (10)$$

The spatial decay of $\mathbf{K}$ depends on the locality of the electronic structure and its off-diagonal matrix elements $K_{i\alpha,j b}$ have nonzero values even when the distance between two atoms $i$ and $j$ is large. However, we only need charge density and Trace($\mathbf{KH}$) (or Trace($\mathbf{K}^{\text{PAO}} \mathbf{H}^{\text{PAO}}$)) to calculate the DFT total energy. In the calculations of Eq. (7) and (10), we need only the **K** (**K**$^{PAO}$) elements which have overlap with two support functions (PAOs) and **H** (**H**$^{PAO}$) with finite range. Therefore, the cutoff regions of **K** and **K**$^{PAO}$ can be set to be the same as **H** (= $2r_{SF}$) and **H**$^{PAO}$ (= $2r_{PAO}$). As a result, Eq. (9) can be also considered as sparse matrix multiplications and can be performed very efficiently on massive parallel computers.[15] Although Eq. (7) and Eq. (10) are equivalent, it is much more efficient to use Eq. (10) to calculate the electronic density, because PAOs are much more localized than multisite support functions and the cutoff range for **K**$^{PAO}$ is shorter than that of **K**.

2.2. Determination of Coefficients in Multisite Support Functions

The accuracy of multisite support functions depends on the accuracy of the coefficients **C**. In principle, **C** should be optimized by minimizing the KS energy of the system numerically with respect to each element in **C**. Since the optimization requires significant effort, we have applied the LFD technique proposed by Rayson et. al.[11, 12] instead of the optimization in this study. The coefficients are determined by projecting the local MOs on some trial vectors in the technique. To calculate the local MOs for the target atom $i$, we construct the subspace of the Hamiltonian and overlap matrices, **H**$_s$ and **S**$_s$, from the matrix elements of **H**$^{PAO}$ and **S**$^{PAO}$ which belong to the target atom $i$ and its neighbor atoms in the region of the radius $r_{LD}$. A local diagonalization is performed by using **H**$_s$ and **S**$_s$, obtaining local MO coefficients **C**$_s$ and eigenvalues $\varepsilon_s$,

$$\mathbf{H}_s \mathbf{C}_s = \varepsilon_s \mathbf{S}_s \mathbf{C}_s. \tag{11}$$

Projecting $\mathbf{C}_s$ on the trial vectors $\mathbf{t}$ and adopting a filter function $f(\varepsilon)$, the contraction coefficients $\mathbf{C}'$ are obtained as

$$\mathbf{C}' = \mathbf{C}_s f(\varepsilon_s) \mathbf{C}_s^T \mathbf{S}_s \mathbf{t}, \qquad (12)$$

where $f(\varepsilon)$ is a Fermi-Dirac function. By setting the chemical potential in $f(\varepsilon)$ to a value around the Fermi level, the effect from unoccupied local MOs with high energy can be eliminated. The trial vectors are arbitrary in principle, and we use the atomic orbitals on the target atoms as trial vectors in this study. Finally $\mathbf{C}'$ is mapped into the corresponding position in the contraction coefficient vectors, which is used as the matrix $\mathbf{C}$ in section 2.1. This procedure is performed for all of the atoms in the system.

In this work, we have introduced double cutoffs and a smoothing function to the LFD method. In the original LFD study, the range of the contraction coefficients is the same as the range of the local diagonalization $r_{LD}$. The use of large $r_{LD}$ generally improves the accuracy of the contraction.[11] However, using CONQUEST, especially in O($N$) calculations, a small cutoff $r_{MS}$ is desirable to reduce the computational cost of sparse-matrix multiplications as shown in Eq. (7), while $r_{LD}$ can be large because the size of the target system is usually much larger than the region of the local diagonalization. Therefore, we study the possibility of using different values for $r_{LD}$ and $r_{MS}$ here: the coefficients are constructed by local diagonalizations including all atoms in $r_{LD}$ range, but only the coefficients from atoms in $r_{MS}$ range are used as $\mathbf{C}$.

The smoothing function is introduced to eliminate the discontinuous change of the accuracy of $\mathbf{C}$ when the number of neighbor atoms varies. The increase or decrease of neighbor atoms in $r_{MS}$ changes the accuracy of $\mathbf{C}$ drastically when $r_{MS}$ is not large. This change will be crucial when we discuss geometrical changes of systems such as energy-volume (E-V) curves and molecular dynamics calculations. Therefore, we have put a smoothing function around $r_{MS}$ to connect the coefficients inside and outside of the region $r_{MS}$ smoothly and avoid sudden changes of accuracy.

Self-consistent-field (SCF) calculations with multisite support functions are performed by two-step procedure in this study: (1) obtain the PAO coefficients of the support functions in Eq. (12) from the given density and (2) perform a SCF calculation without changing these coefficients. The coefficients are then updated after self-consistency has been reached, and we then repeat (1) and (2) until the energy is converged to within a certain threshold. The determination of the coefficients without optimizations may cause uncertainties in the SCF and force calculations. By the simple projection on the trial vectors, the coefficients and the electron density cannot be determined to minimize the total energy at the same time and the Pulay force from the coefficients, which is reported to be negligible for large $r_{MS}$,[11] cannot be calculated analytically. The SCF procedure in the present study converges in most cases although it is not rigorously variational. These problems will be serious when $r_{MS}$ is too small; we discuss them further in the Appendix. The optimization of the coefficients in future work will solve these problems.

3. RESULTS

In this section, we report benchmark calculations on a simple semiconductor (bulk Si in Sec. 3-1), a metal (bulk Al in Sec. 3-2) and a large-scale organic insulator (hydrated DNA in Sec. 3-3), to show the accuracy and efficiency of multisite support functions. For the first two systems, we first investigate the energy convergence on the cutoff range $r_{MS}$. Then, we mainly discuss the bulk modulus $B_0$ and the optimized lattice constant $a_0$, since we do not need the absolute convergence of the total energy in many cases, but only need the energy difference in actual research. We also clarify the accuracy of the band structure or the density of states near the Fermi level, including the unoccupied states, although the support functions are mainly intended to represent the occupied wave functions or density matrix. For the DNA system, the reduction of the computational time is also discussed.

As for the calculation conditions, we use the local density approximation (LDA)[17] and TZP PAO basis sets for the first two bulk systems, while the generalized gradient approximation (GGA)[18] and DZP are adopted for the DNA system. In the generation of DZP basis sets, we mainly used the method of the energy shift (100meV for DNA and 300meV for bulk Si and Al) implemented in Siesta.[7] The actual cutoffs of the PAOs will be given in each section. The cutoff energy of the charge density grid is 80 Hartree for bulk Si and Al, and 75 Hartree for the DNA system. The bulk modulus $B_0$, the equilibrium volume $V_0$ and the lattice constant $a_0$ are calculated by fitting E-V curves with Birch-Murnaghan equation. As for the filter function $f(\varepsilon)$ in Eq. (12), we adopted the Fermi-Dirac function with $kT = 0.1$. As explained in section 2.2, the chemical potential in $f(\varepsilon)$ is set to the mean value of the energies of the highest occupied molecular orbital (HOMO) and the lowest unoccupied molecular orbital (LUMO) for each subspace.

Note that we use exact diagonalization, not O($N$) method in this work, because we would like to concentrate on the accuracy and efficiency of the multisite support functions. The number of k-points used in the diagonalization method are (4, 4, 4) and (6, 6, 6) for Si and Al, respectively, and only the Γ point is considered for the DNA system. All of the results in this section are calculated using the SCF procedure described in section 2.2. The effect of the update of the coefficients in the multisite support functions in the present SCF procedure is discussed in Appendix.

3.1. Bulk Si

We first consider crystalline silicon and investigate the energy convergence with respect to the multisite range $r_{MS}$. The TZP PAOs used here have 17 PAOs per Si atom; three $s$, three $p$ and a $d$ PAOs.[19] These 17 PAOs are contracted into four multisite support functions. When calculating coefficients in multisite support functions using Eq. (12), we have tried two sets of trial vectors, the most delocalized $s$ and $p$ PAOs (TVEC1) with $r_{PAO}$ = 6.6 bohr and the most localized ones (TVEC3) with $r_{PAO}$ = 3.5 bohr. The difference between the DFT total energies per atom by primitive TZP and those by multisite support functions are shown in Fig.1 (actual values are available in the supporting information). The cutoff of the local diagonalization $r_{LD}$ is set to be equal to $r_{MS}$, and $r_{MS}$ is changed from 0.5 to 26.0 bohr. When $r_{MS}$ is set to 0.5 bohr, only the target atom is included in the multisite region. Fig. 1 illustrates the smooth energy-convergence with respect to $r_{MS}$. The energy calculated with $r_{MS} \geq 8.0$ bohr converges to that by using the primitive support functions within the order of $10^{-2}$ eV, implying that only the atoms up to second neighbor are required to achieve mHartree accuracy. We also observe that using the more localized TVEC3 results in fast convergence and a little higher energy in convergence. However, the difference is less than 1mHartree when $r_{MS}$ > 8.0 bohr and we show the results with TVEC3 hereafter. We have confirmed that the results with TVEC1 are essentially the same; though there is a flattening off of the curve for TVEC3, this may because TVEC3 is quite localized but this is a negligible effect.

We now turn to the effect of varying $r_{LD}$, considering the E-V curves shown in Fig. 2. Multisite support functions with several combinations of $r_{LD}$ and $r_{MS}$ are investigated, which are denoted as ($r_{LD}$-$r_{MS}$): (17.0-17.0), (8.0-8.0) and (5.0-5.0) in Fig. 2 (a) in which $r_{LD}$ is set to be equal to $r_{MS}$, and (17.0-8.0) and (17.0-5.0) in Fig. 2(b) in which $r_{LD}$ is set to be larger than $r_{MS}$. The multisite region with $r_{MS}$ of 8.0 and 5.0 bohr contains up to second- and the nearest neighbor atoms, respectively. Multisite support functions with the smoothing (SMTH) functions are also examined by comparing (17.0-11.0) and (17.0-11.0)$_{SMTH}$ in Fig. 2(c). The results with primitive support functions are also shown in the figure. Total energy and volume per unit cell are plotted in vertical and horizontal axes, respectively. The equilibrium volume $V_0$ is also shown in the figure. The bulk modulus $B_0$ and optimized lattice parameter $a_0$ are listed in Table 1.

In Fig. 2(a), the E-V curve of (17.0-17.0) is found to be close to the one of primitive support functions, the difference of the energy between them being less than 0.005 eV/atom for all volumes. The result using (8.0-8.0) shows a higher energy by about 0.03 eV/atom, but the energy difference from that of primitive functions is almost the same for different volumes. As a result, the deviations of $B_0$ and $a_0$ of (8.0-8.0) from those of primitive functions are very small, less than 1 % for $B_0$ and 0.05 % for $a_0$, as shown in Table 1. We can conclude that the accuracy of (17.0-17.0) and (8.0-8.0) are almost comparable, meaning that the multisite support functions are converged with the cutoffs of 8.0 bohr including up to second-neighbor atoms. On the other hand, the energy curve in the (5.0-5.0) case, in which the multisite region includes only the nearest neighbors, is higher in energy than the result of the primitive ones by about 0.2 eV/atom and the difference changes when the volume varies. As a result, the deviations of $B_0$ and $a_0$ are both about ten-times larger than (8.0-8.0). The error of $a_0$, 0.7 %, may be still acceptable in many cases, but that of $B_0$, about 10.1 %, may cause a serious problem in some cases. (Note that the typical error for the bulk modulus by LDA is larger than 10 %.)

As explained in Sec. 2.2, using a smaller $r_{MS}$ than $r_{LD}$ is important to reduce the computational cost, especially in O($N$) calculations. The results using larger $r_{LD}$ than $r_{MS}$ are shown in Fig. 2(b). In the case of bulk Si, we can see that the accuracy of E-V curve is not improved even if we use larger $r_{LD}$ than $r_{MS}$. In the case of $r_{MS}$ being 5.0 bohr, the energy becomes even higher. The accuracy of $B_0$ can be improved in some cases ((17.0-5.0), (17.0-8.0)), but we cannot see clear tendency; in general $r_{MS}$ seems to determine the properties. Since the convergence with respect to $r_{LD}$ is very fast in this system, it may be difficult to see the effect of having larger $r_{LD}$ in the E-V curve.

In Fig. 2(c), a discontinuity is found in the E-V curve of (17.0-11.0) case. This is due to the change of the number of the neighbor atoms in the multisite region ($r_{MS}$ = 11.0 bohr) when the volume of the unit cell changes. The accuracy of the calculated energy with the multisite support functions depends on the number of the neighbor atoms significantly when $r_{MS}$ is small. Thus, we need a smoothing function to get a smooth E-V curve. Here, we use Fermi-Dirac function as a smoothing function, whose center position $r_{SMTH}$ is set to be 1 bohr smaller than $r_{MS}$, and the width of the function is set to 0.5 bohr. With this procedure, the curve of (17.0-11.0)$_{SMTH}$ is obtained as a smooth E-V curve and the sudden change of the energy due to the change of the number of neighbor atoms is eliminated.

We next compare the band structures obtained with primitive and multisite support functions in Fig. 3. Three kinds of multisite support functions are investigated here: (17.0-17.0), (17.0-8.0) and (8.0-8.0). The band structure calculated with the primitive TZP functions is close to that by plane-wave calculations.[21,22] All of the three multisite support functions reproduce the occupied band structure with high accuracy, including the indirect band gap of Si. On the other hand, the unoccupied band structures with (17.0-17.0) and (17.0-8.0) are close to that with the primitive support functions, while that with (8.0-8.0) is clearly different from them. This result suggests that the accuracy of unoccupied band structures can be improved by increasing $r_{LD}$, although the multisite support functions are constructed based mostly on the occupied local orbitals. From this result, we expect that the combination of small $r_{MS}$ and large $r_{LD}$ is effective even for bulk Si systems.

To assess the basis set dependence, TZDP and DZP PAOs were also employed for bulk Si to assess the basis set dependence. This assessment confirms that we can construct multisite support functions from various PAOs having a good accuracy. The details of the calculations are presented in the supporting information.

We also examine the forces calculated with the multisite support functions. As mentioned in section 2.2, the coefficients **C** depend on the atomic positions but are not optimized numerically so that Pulay forces from the change of the coefficients remain in principle. However, since the accuracy of the coefficients by the present projection method is rather high, we simply neglect the term and calculate Pulay forces by considering only the first derivative of the primitive PAOs with respect to the atomic positions in this study. The accuracy of the forces calculated in this way is checked for the system in which a Si atom in bulk Si (8 atom in an unit cell) is displaced by 1.0 bohr to the nearest neighbor atom. The total force of the shifted atom is calculated as 0.1767, 0.1765 and 0.1768 hartree/bohr by primitive TZP PAOs and multisite support functions (17.0-17.0) and (8.0-8.0), respectively. The deviation of the force calculated by multisite support functions is only about 0.2 mhartree/bohr even if the term from the coefficients is neglected.

The variation of total energy and the force on the shifted Si atom are investigated by changing the displacement of the atom. ($r_{LD}$-$r_{MS}$) are set to (17.0-7.5) bohr. The results with and without smoothing functions are shown in Fig. 4. In Fig. 4(a), a sudden change of the total energy is found around a displacement of 0.3 bohr in the curve for the calculations without using smoothing functions. The discontinuity in the total energy is due to the change of the neighbor atoms and is eliminated by using the smoothing function. On the other hand in Fig 4(b), both calculations with and without the smoothing function provide almost the same forces without any sudden changes. The inconsistency between the energy and forces in the calculations without smoothing functions can cause serious problems when we perform geometry optimizations or molecular dynamics simulations. Thus, the use of smoothing functions is important for reliable molecular dynamics or structure relaxations.

3.2. Bulk Al

In this section, we show the results of metallic Al system. 17 TZP PAOs[23] are contracted into four multisite support functions, and r$_{PAO}$ of ($s$, $p$) in TVEC1 and TVEC3 are (6.1, 7.1) and (5.0, 5.5) bohr, respectively. Figure 5 shows the energy difference from the DFT energy per atom of bulk Al using primitive TZP with respect to $r_{MS}$ (actual values are in the supporting information).

Similarly to the case of Si, the calculated energy is converged in mHartree order when $r_{MS} \geq 8.0$ bohr and the multisite region includes the atoms up to the second neighbors. Although the locality of the electronic structure is very different between metallic Al and semiconducting Si systems, the present convergence behavior is only for the support functions, and is not significantly different from that of bulk Si, if perhaps a little slower.

We next investigate the E-V curves for various multisite support functions, shown in Fig. 6, which shows the results with primitive and five sets of multisite support functions, (16.0-16.0), (8.0-8.0), (6.0-6.0), (16.0-8.0) and (16.0-6.0). Table 2 lists the $B_0$ and $a_0$ values of bulk Al calculated from the E-V curves. The curve of (16.0-16.0) is very close to that of primitive support functions. The curve of (8.0-8.0) provides $B_0$ and $a_0$ with reasonable accuracy, but the deviations (3.39 % for $B_0$ and 0.51 % for $a_0$) from those of the primitive TZP PAOs are larger than the case of bulk Si having the same cutoffs. The deviations of $B_0$ and $a_0$ using (6.0-6.0) are 13.6 % and 1.34 %, which are large and not negligible. However, if we use larger $r_{LD}$, keeping the same cutoff $r_{MS}$, these deviations become much smaller: 2.41 % for $B_0$ and 0.40 % for $a_0$ in the case of (16.0-8.0), and 3.25 % and 0.57 % for (16.0-6.0) case. The deviations of the total energy are also reduced. From these results, we can conclude that the combination of small $r_{MS}$ and large $r_{LD}$ is an effective method to keep certain accuracy with small $r_{MS}$.

Figure 7 shows the band structure of bulk Al calculated with primitive and multisite support functions (16.0-16.0), (16.0-8.0) and (8.0-8.0). Similarly to the case of Si, the occupied band structures obtained by multisite support functions are close to that by primitive support functions. However for bulk Al, the numbers of unoccupied band structures in primitive and multisite support functions are different. Since the number of multisite support functions for an Al atom is four, only four bands are obtained in the Γ→X region. Note that the highest band in Fig. 7 (b) –

(c) is doubly degenerate and the corresponding band in Fig. 7(a) (the third one from the bottom) is also doubly degenerate. Comparing the existing unoccupied bands, we can again say that the increase of $r_{LD}$ improves the accuracy of energy levels in the unoccupied region. Especially the energy levels of (16.0-16.0) show a good agreement.

3.3. Hydrated DNA

As a next target, we examine the DNA system in water solvent, to investigate the accuracy of multisite support functions for organic insulators. The system is B-DNA decamer 5'-d(CCATTAATGG)2-3', which contains 634 DNA atoms with 932 hydrating water molecules and 9 Mg counter ions, totalling 3439 atoms. The structure of the hydrated DNA system is shown in the previous paper.[24] In the present calculations, 27883 primitive DZP PAOs are contracted into 7447 multisite support functions (8.0-8.0) and (16.0-16.0).

It should be noted that we have observed an instability in the update of the PAO coefficients of multisite support functions with the SCF procedure, in the case of (8.0-8.0). When the update procedure was close to convergence, small energy fluctuation ($10^{-6}$ hartree per atom) was observed and we stopped the procedure when the energy was at a minimum. We show here the result obtained by this procedure for the case of (8.0-8.0). This instability is probably because the present method is not variational as explained in section 2.2 and discussed in the Appendix, and should be removed if the PAO coefficients are determined by optimization. It should be also noted that this instability is not found when $r_{MS}$ is large.

Table 3 shows that the HOMO-LUMO gaps by the two sets of multisite support functions are both very close to that by primitive DZP: the deviations are about only 0.01 eV. As is well known in the GGA calculations, the calculated HOMO-LUMO gaps are smaller than the typical HOMO-LUMO gaps reported for DNA systems, being about 3 – 4 eV.[25]

Figure 8 shows the DOS by SCF calculations obtained by using a Gaussian broadening with the half-width of 0.1 eV. The Fermi level is set to be origin in the figure. The results by (8.0-8.0) and (16.0-16.0) are both similar to that by primitive DZP in the occupied region, as in the case of bulk Si and Al. For the unoccupied region, the DOS of (16.0-16.0) is much closer to that of DZP than (8.0-8.0), which indicates that the larger cutoffs improve the accuracy of the unoccupied energy levels.

Finally, we compare the computational time spent for one SCF iteration, which includes the times for the construction of matrices and one-shot diagonalization, to examine the efficiency of the multisite method. We used 96 cores (16nodes, 6 cores per node) of the supercomputer system HITACHI HA8000-tc for each calculation. Using the multisite support functions reduces the computational time significantly: The times for DZP, (8.0-8.0) and (16.0-16.0) are 13342, 889, and 2042 seconds, respectively. Note that the data for the case of multi-site support functions, (8.0-8.0) and (16.0-16.0), include the times for constructing multisite support functions. The difference between (8.0-8.0) and (16.0-16.0) mainly comes from the time to construct the multi-site support functions, but in both cases the computation time is much smaller than the case using primitive DZP. Therefore, we conclude that the use of multisite support functions is an efficient and promising method to investigate the electronic structures of large systems with reasonable computational costs without losing accuracy.

## 4. CONCLUSIONS

The multisite support functions which consist of not only the PAOs on the target atom but also those on neighbouring atoms have been developed in CONQUEST, a DFT code for large-scale systems. The PAO coefficients of the multisite support functions are calculated from the local molecular orbitals around the target atom using the localized filter diagonalization method.[11, 12] In this work, we have newly introduced the double cutoff method in which two different values are used: the cutoff for the local diagonalization $r_{LD}$, and that of the multisite support functions $r_{MS}$. A smoothing function has been also introduced to make the energy change smooth even when the number of neighbor atoms in the multisite region changes.

The accuracy and efficiency of the multisite support functions have been examined for bulk Si, bulk Al and large-scale hydrated DNA systems. For the first two systems, we have confirmed that even if the number of multisite support functions is reduced to be same as the minimal PAO basis set (SZ), the accuracy is almost the same as the primitive TZP PAO basis functions. The energy convergence with respect to the cutoff $r_{MS}$ shows we can achieve enough accuracy if the multisite region includes up to second neighbor atoms of the target atom. Furthermore, the multisite support functions can, even when its cutoff is rather small, provide the bulk modulus and lattice constant in comparable precision with primitive support functions. The tendency of semiconducting Si and metallic Al are almost the same. The use of smoothing functions can eliminate the sudden change of the accuracy according with the number of neighbor atoms and help to keep the consistency between energies and forces. The band structures of bulk Si and Al, and the density of states of a hydrated DNA system calculated by multisite support functions are also accurate in the occupied region even if we use small cutoffs. Large $r_{LD}$ helps to yield not only occupied but also unoccupied band structures accurately. It has been demonstrated that the combination of small $r_{MS}$ and large $r_{LD}$ by the double cutoff method improve the accuracy of energy levels as well as the structural parameters. Thus, we expect that multisite support functions can be a powerful tool also for the excitation energies of large-scale systems if we use them in time-dependent DFT calculations, whose accuracy depend on the orbital energies largely. We will report on the implementation of TDDFT in CONQUEST in a forthcoming paper[26].

We have also reported the significant reduction of the computational time by the present multisite method with CONQUEST, in the SCF calculations on a large-scale DNA system containing about 3400 atoms. The reductions of the number of support functions with the multisite method should be efficient not only in the conventional diagonalization method but also in O($N$) calculations because the computational cost scales cubically to the number of support functions on each atom for both calculations.

Appendix

In this appendix, we discuss the problem of the SCF procedure with the LFD method and investigate the effect of the update of the PAO coefficients of the multisite support functions.

Let us first discuss the possible problem in the update of the PAO coefficients with the SCF procedure, explained in Sec. 2.2. If we fix the support functions, we can obtain the SCF charge density without any problems (procedure (2) in Sec. 2.2). On the other hand, if the charge density is given, the DFT Hamiltonian is calculated from the density and the PAO coefficients of multisite support functions are obtained using the LFD method (procedure (1) in Sec. 2.2). If the accuracy of the multisite support functions is the same as that of the primitive support functions, it is guaranteed that there exists a set of PAO coefficients which provides the SCF charge density giving the Hamiltonian consistent with the PAO coefficients. Or, if we calculate the PAO coefficients which *minimize* the DFT total energy calculated with the SCF charge density given by the support functions, there will be no problems to calculate the support functions and SCF charge density simultaneously. However, if the accuracy of the multisite support functions is not sufficient, for instance if the PAO coefficients are calculated simply by the projection method, there is no guarantee that we can obtain consistent multisite support functions and SCF charge density. In fact, as we have pointed out in Sec. 3.3, we have observed an instability of the two step procedure, explained in Sec. 2.2, to calculate the multisite support functions with SCF.

One possible way to solve this problem is to simply use the PAO coefficients calculated from the initial charge density without updating the coefficients. Although this scheme sounds a little crude, it may be very effective and reasonably accurate for practical purposes, since we expect that the support functions are only a kind of basis set and may not change significantly while repeating procedures (1) and (2). In this respect, we have investigated the accuracy of the method, i.e., one-shot procedure of (1) and (2) for bulk Si and the hydrated DNA systems. The initial charge densities were the superposition of the atomic charge densities. Table A1 shows the $B_0$ and $a_0$ values of Si calculated with this one-shot procedure. By comparing Table A1 with Table 2, we find that the deviations of the results with the initial coefficients and those with the updated coefficients are small, less than 1 % for $B_0$ and 0.2 % for $a_0$. These small deviations indicate that the initial and updated coefficients are not significantly different for bulk Si. On the other hand, the electronic structure calculated by the initial and updated coefficients are found to be different in the hydrated DNA system. Table A2 shows the HOMO-LUMO gaps of the hydrated DNA system calculated with the multisite support functions with the initial coefficients. Comparing Table A2 and Table 3, the deviations of the HOMO-LUMO gaps with the initial and updated coefficients change by more than 0.5 eV. The DOS obtained with the multisite support functions (16.0-16.0) with the initial coefficients in Fig. A1 also shows a large difference to that with the updated coefficients in Fig. 8. These differences indicate that the changes of the coefficients are large during the update procedure. Note that the system includes various charged groups, such as Mg or $PO_4$, although all of them were simply considered as neutral in constructing the initial charge density.

As can be seen in the example of the hydrated DNA case, we need to use this simple method carefully, but the method is reasonably accurate in some cases. Even in the hydrated DNA system, if we calculate the initial charge density during MD from the density matrix in the previous step of MD with SCF, the deviations are expected to be much smaller. Furthermore, if we follow the optimization procedure after the one-shot LFD method, the initial coefficients should be very close to the optimized ones. The development towards this direction is now under way and we believe the combination of LFD and optimization methods is promising for the efficient and accurate calculations.

In the end of this appendix, we would like to make a short comment. CONQUEST can employ non-self-consistent (NSC) calculations using the Harris-Foulkes energy functional, as well as the self-consistent calculations. The accuracy and the efficiency of the NSC method are shown in our previous papers,[27–29] and there should be no problems of the LDF method discussed in this section with the NSC method. We expect that the method is efficient and robust in many cases of the actual research.

FIGURES. (All figures are in actual size.)

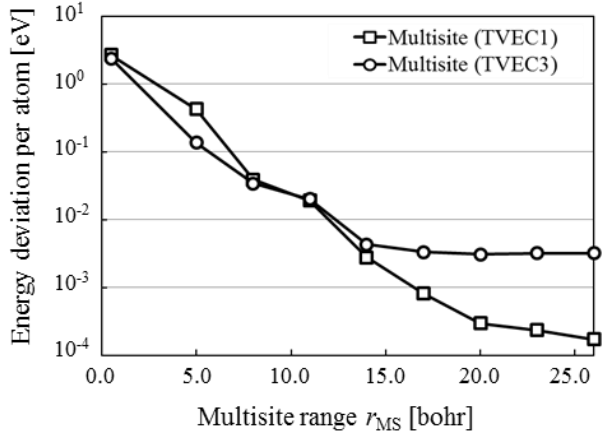

**Figure 1.** Deviation of total energy per atom [eV] of bulk Si with multisite support functions from the energy with primitive support functions with respect to the multisite range $r_{MS}$. Local diagonalization range $r_{LD}$ is equal to $r_{MS}$. The squares and circles correspond to the results by multisite support functions with the most delocalized (TVEC1) and the most localized (TVEC3) trial vectors.

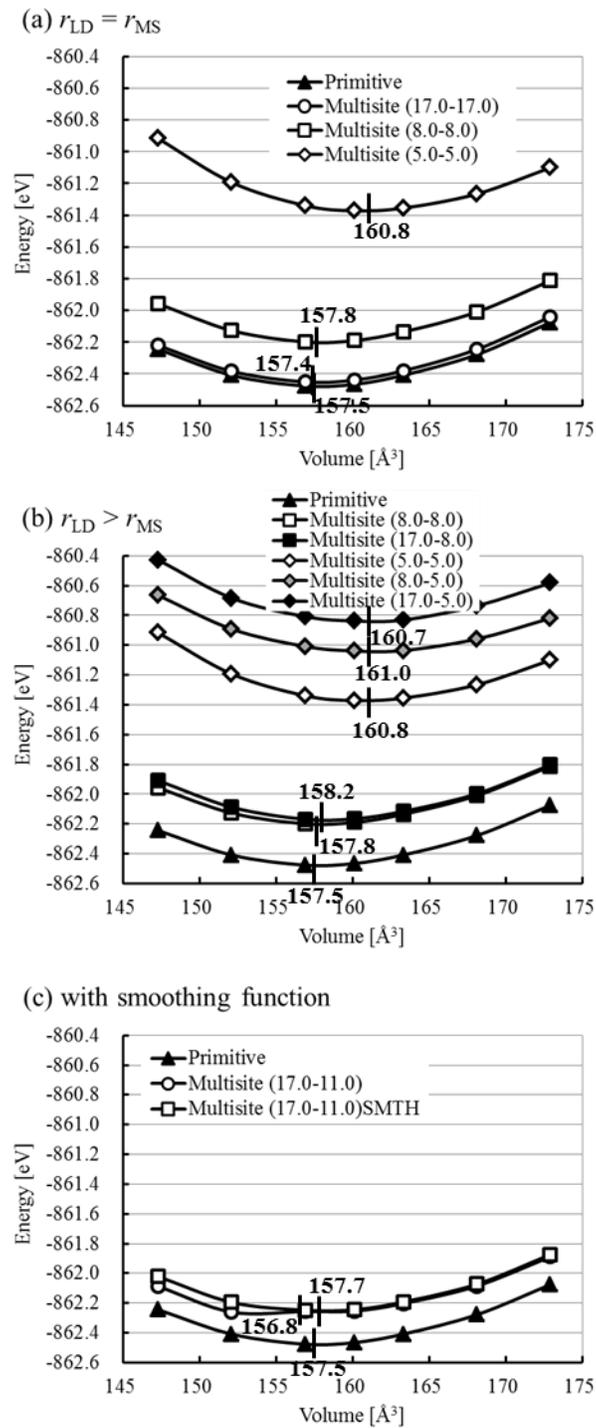

**Figure 2.** Energy-Volume curves of bulk Si. The triangles correspond to total energies of unit cells with primitive support functions, and circles, squares and diamonds correspond to those with multisite support functions. $r_{LD}$ and $r_{MS}$ in multisite support functions are in parentheses as ($r_{LD}$-$r_{MS}$). Three types of multisite support functions are adopted: (a) $r_{LD} = r_{MS}$, (b) $r_{LD} > r_{MS}$ and (c) with smoothing functions. The lowest energy volume for each method is shown as a vertical line with a number, while experimental value is 160.2 Å$^3$.

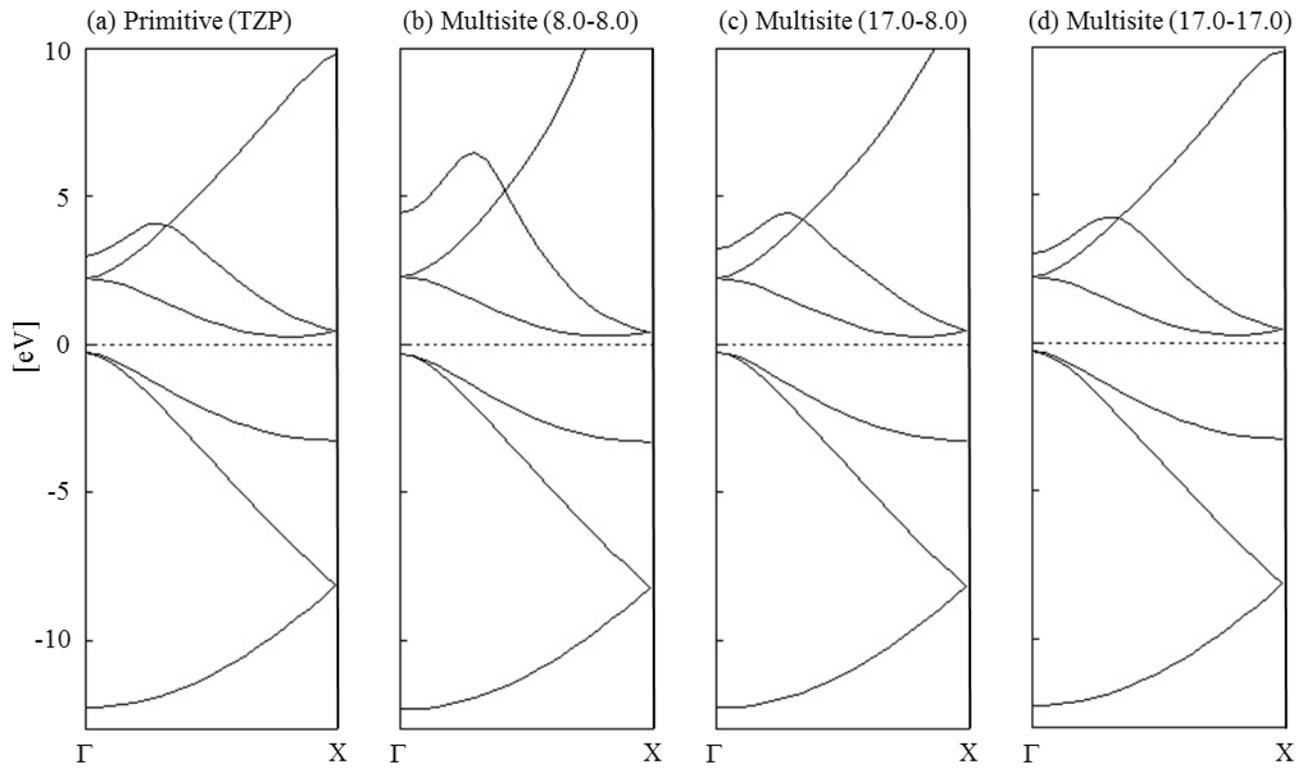

**Figure 3.** Band structures of bulk Si obtained with primitive and multisite support functions. The structure between $\Gamma$ and X points is shown. $r_{LD}$ and $r_{MS}$ in multisite support functions are in parentheses as ($r_{LD}$-$r_{MS}$). The Fermi level is set to be the energy zero.

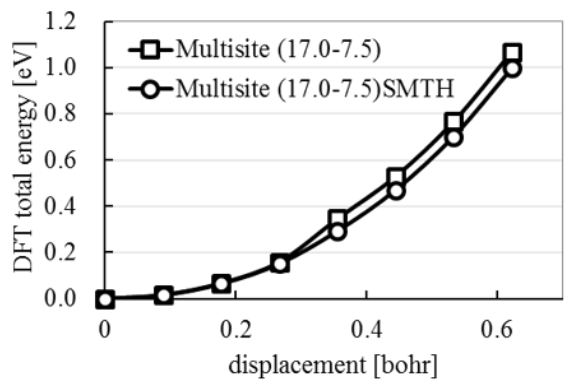

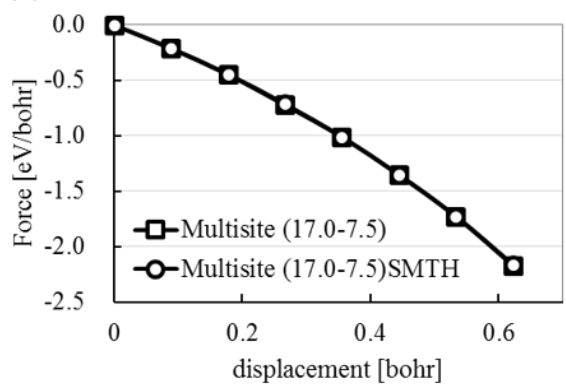

**Figure 4.** Variation of (a) DFT total energies and (b) forces with respect to the displacement of a Si atom in bulk Si. Multisite support functions with and without smoothing functions (SMTH) with $(r_{LD}\text{-}r_{MS}) = (17.0\text{-}7.5)$ bohr are used.

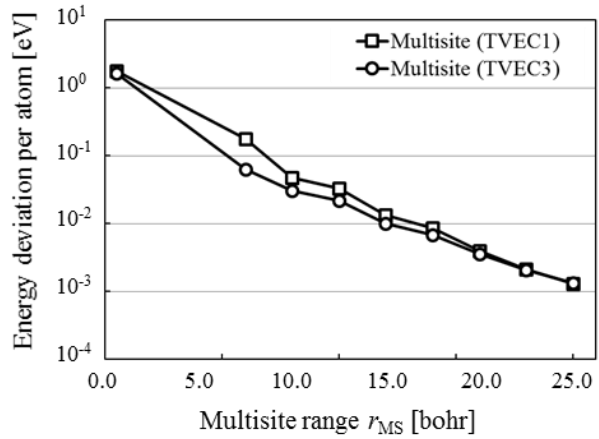

**Figure 5.** Deviation of total energy per atom [eV] of bulk Al with multisite support functions from the energy with primitive support functions with respect to the multisite range $r_{MS}$. Local diagonalization range $r_{LD}$ is equal to $r_{MS}$. The squares and circles correspond to the results by multisite support functions with the most delocalized (TVEC1) and the most localized (TVEC3) trial vectors.

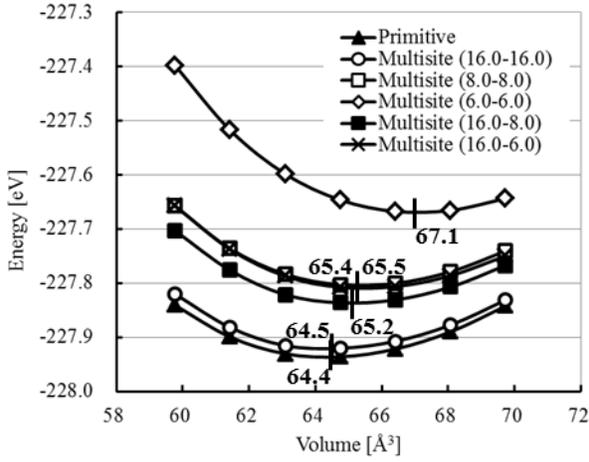

**Figure 6.** Energy-Volume curves of bulk Al. The triangles correspond to total energies of unit cells with primitive support functions, and circles, squares, diamonds and crosses correspond to those with multisite support functions. $r_{LD}$ and $r_{MS}$ in multisite support functions are in parentheses as ($r_{LD}$-$r_{MS}$). The minimal volume for each method is shown as a vertical line with a number, while experimental value is 66.4 Å$^3$.

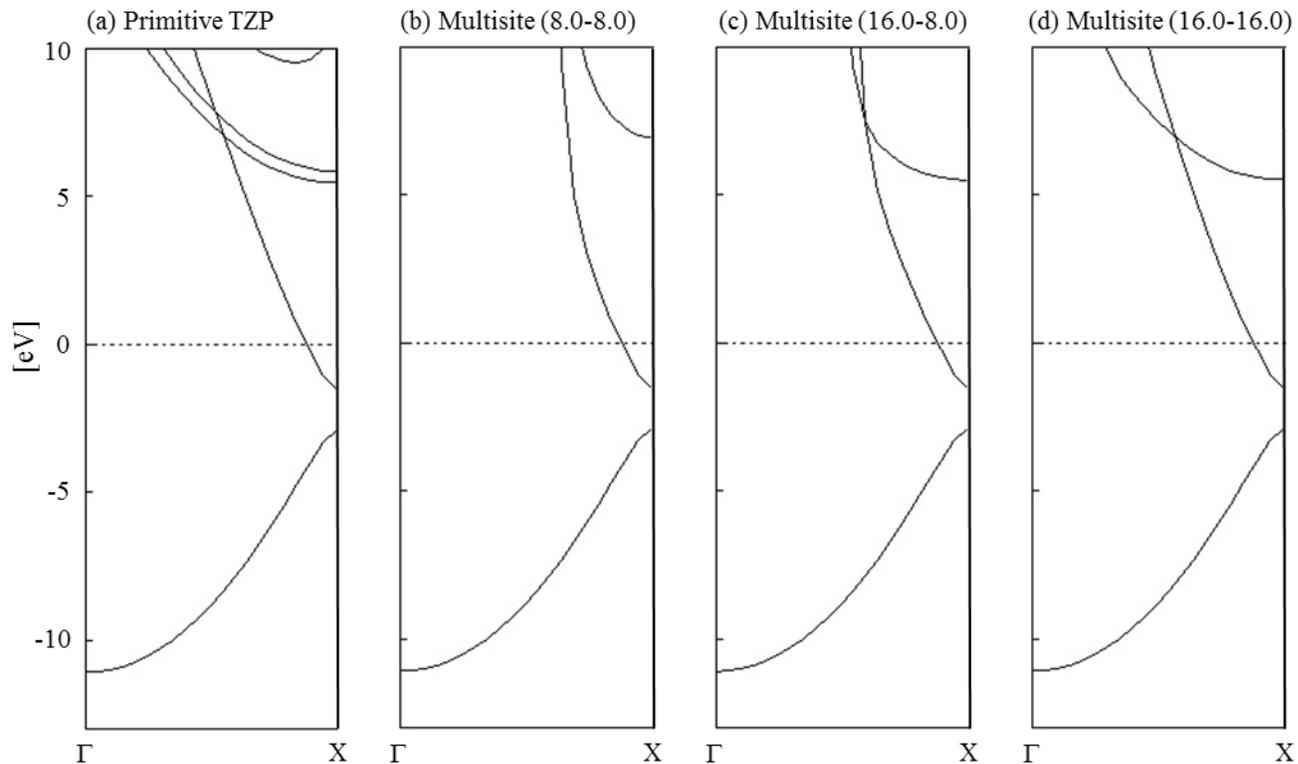

**Figure 7.** Band structures of bulk Al obtained with primitive and multisite support functions. The structure between Γ and X points is shown. $r_{LD}$ and $r_{MS}$ in multisite support functions are in parentheses as ($r_{LD}$-$r_{MS}$). The Fermi level is set to be the energy zero.

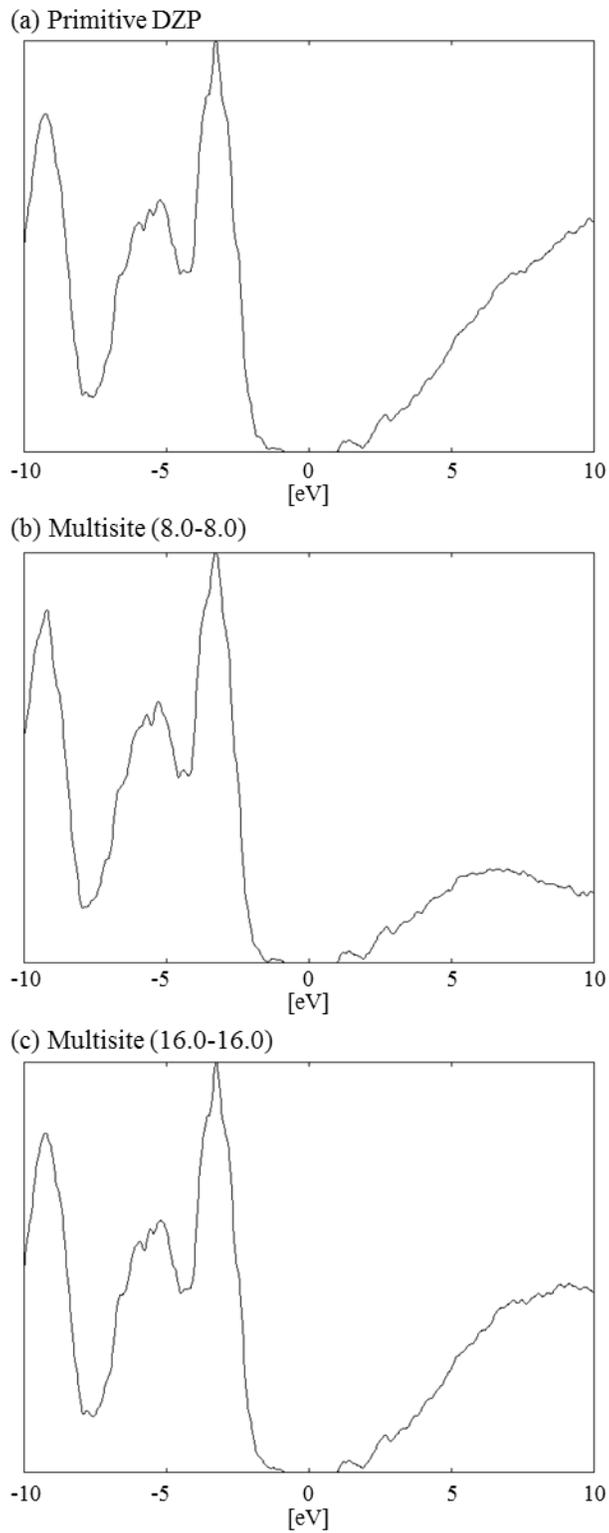

**Figure 8.** Density of states of the 3439-atom hydrated DNA calculated with (a) primitive DZP, multisite (b) (8.0-8.0) and (c) (16.0-16.0) support functions. $r_{LD}$ and $r_{MS}$ in multisite support functions are in parentheses as ($r_{LD}$-$r_{MS}$). The Fermi level is set to be the energy zero.

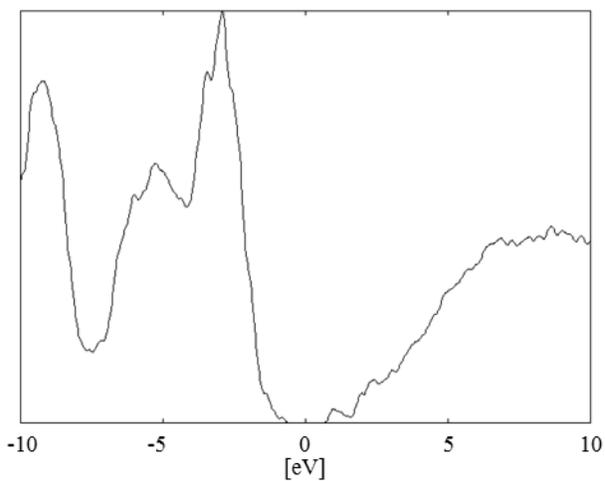

**Figure A1.** Density of states of the 3439-atom hydrated DNA calculated with multisite support functions ($r_{LD}$-$r_{MS}$) = (16.0-16.0) constructed from the initial charge density. The Fermi level is set to be the energy zero.

TABLES.

**Table 1.** Bulk modulus and lattice constants of bulk Si calculated with TZP-primitive and multisite support functions. Percent deviation from the results by primitive support functions are shown in parentheses.

|  | $B_0$ [Gpa] | $a_0$ [Å] |
|---|---|---|
| Primitive | 99.5 | 5.401 |
| Multisite ($r_{LD}$ - $r_{MS}$) | | |
| (17.0-17.0) | 100.7 (1.12) | 5.400 (0.02) |
| (8.0-8.0) | 98.8 (0.70) | 5.404 (0.05) |
| (5.0-5.0) | 109.6 (10.08) | 5.439 (0.70) |
| (17.0-8.0) | 99.2 (0.40) | 5.408 (0.14) |
| (8.0-5.0) | 89.8 (9.76) | 5.440 (0.72) |
| (17.0-5.0) | 101.6 (2.07) | 5.437 (0.67) |
| (17.0-11.0) | 88.0 (11.60) | 5.393 (0.15) |
| (17.0-11.0)$_{SMTH}$ | 96.2 (3.40) | 5.403 (0.03) |
| Exptl. | 100 | 5.431 |

[a] Reference 20.

**Table 2.** Bulk modulus and lattice constants of bulk Al calculated from total energies with TZP-primitive and multisite support functions. Percent deviation from the results by primitive support functions are shown in parentheses.

|  | $B_0$ [Gpa] | $a_0$ [Å] |
|---|---|---|
| Primitive | 79.8 | 4.009 |
| Multisite ($r_{LD}$ - $r_{MS}$) | | |
| (16.0-16.0) | 80.0  (0.19) | 4.011  (0.06) |
| (8.0-8.0) | 82.5  (3.39) | 4.030  (0.51) |
| (6.0-6.0) | 90.7  (13.62) | 4.063  (1.34) |
| (16.0-8.0) | 81.7  (2.41) | 4.025  (0.40) |
| (16.0-6.0) | 82.4  (3.25) | 4.032  (0.57) |
| Exptl. | 76 | 4.050 |

[a] Reference 20.

**Table 3.** HOMO-LUMO gap of the hydrated DNA system [eV] calculated with primitive DZP and multisite support functions.

|  | Gap [eV] |
|---|---|
| Primitive DZP | 2.17 |
| Multisite ($r_{LD}$ - $r_{MS}$) |  |
| (16.0-16.0) | 2.17 |
| (8.0-8.0) | 2.16 |

[a] Multisite support functions with the initial coefficients.

[b] Multisite support functions with the updated coefficients.

**Table A1.** Bulk modulus and lattice constants of bulk Si calculated with multisite support functions with the initial linear-combination coefficients. Percent deviation from the results by primitive support functions are shown in parentheses.

|  | $B_0$ [Gpa] | $a_0$ [Å] |
|---|---|---|
| Primitive | 99.5 | 5.401 |
| Multisite ($r_{LD}$ - $r_{MS}$) |  |  |
| (17.0-17.0) | 100.0 (0.40) | 5.406 (0.09) |
| (8.0-8.0) | 99.5 (0.08) | 5.410 (0.18) |

**Table A2.** HOMO-LUMO gap of the hydrated DNA system [eV] calculated with primitive DZP and multisite support functions with the initial linear-combination coefficients.

|                                | Gap [eV] |
|--------------------------------|----------|
| Multisite ($r_{LD}$ - $r_{MS}$) |          |
| (16.0-16.0)                    | 1.63     |
| (8.0-8.0)                      | 1.08     |

## ASSOCIATED CONTENT

Tables of the deviations of the DFT energies of multisite support functions from those of the primitive TZP PAOs for bulk Si and Al; Tables of $B_0$ and $a_0$ calculated with multisite support functions consisting of DZP and TZDP PAOs. This material is available free of charge via the Internet at http://pubs.acs.org."

## AUTHOR INFORMATION

**Corresponding Authors**

*E-mail: NAKATA.Ayako@nims.go.jp

**E-mail: MIYAZAKI.Tsuyoshi@nims.go.jp

## ACKNOWLEDGMENT

A.N. is funded by ICYS-NAMIKI, NIMS. This work is supported by KAKENHI projects by MEXT (No. 22104005) and by JSPS (No. 25810015 and 26610120), Japan. This work is also supported by the Strategic Programs for Innovative Research (SPIRE) and the Computational Materials Science Initiative (CMSI), Japan. Calculations where performed in the Numerical Materials Simulator at NIMS, Tsukuba, Japan and the supercomputer HA8000 system at Kyushu-university, Fukuoka, Japan.

For Table of Contents use only

(in actual size)

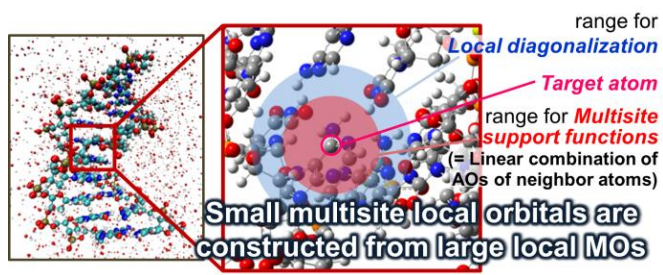

Title of the paper:

Efficient Calculations with Multisite Local Orbitals in a large-scale DFT Code CONQUEST

Authors: Ayako Nakat, David R. Bowler, Tsuyoshi Miyazaki